\documentclass[letter]{aa} 
\usepackage{epsfig}
\usepackage{amsmath}
\usepackage{subfigure}
\usepackage{natbib}
\usepackage{multirow}
\usepackage{color}
\usepackage{longtable}
\usepackage{graphicx}
\usepackage{epstopdf}
\usepackage{txfonts}

\begin{document}

\title{Discovery of the first Ca-bearing molecule in space: CaNC
\thanks{This work was based on observations carried out with the 
IRAM 30-meter telescope. IRAM is supported by INSU/CNRS (France), 
MPG (Germany) and IGN (Spain)}}

   \author{
   J.~Cernicharo\inst{1}\thanks{corresponding author. email: jose.cernicharo@csic.es}, 
   L.~Velilla-Prieto\inst{1,2},
   M.~Ag\'undez\inst{1},
   J.R.~Pardo\inst{1},
   J.P.~Fonfr\'ia\inst{1},
   G.~Quintana-Lacaci\inst{1},
   C.~Cabezas\inst{1},
   C.~Berm\'udez\inst{1},
   M.~Gu\'elin\inst{3}
}

\institute{Grupo de Astrof\'isica Molecular. Instituto de F\'isica Fundamental (IFF-CSIC).
C/Serrano 121, 28006 Madrid, Spain
\and Department of Space, Earth and Environment, Chalmers University of Technology, Onsala Space Observatory, 439 92 Onsala, Sweden
\and 
Institut de Radioastronomie Millim\'etrique, 300 rue de la Piscine, F-38406, 
Saint Martin d'H\`eres, France
}

   \date{7 June 2019 / 14 June 2019}

\abstract{
We report on the detection of calcium isocyanide, CaNC, in the
carbon-rich evolved star IRC+10216. We derived a column density for this species of 
(2$\pm$0.5)$\times$10$^{11}$\,cm$^{-2}$. Based on the observed 
line profiles and the modelling of its emission through the envelope,
the molecule has to be produced in the intermediate and outer layers of the circumstellar envelope
where other metal-isocyanides 
have previously been found in this source. The abundance ratio of CaNC relative to MgNC and FeCN is 
$\simeq$1/60 and $\simeq$1, respectively.
We searched for the species CaF, CaCl, CaC, CaCCH, and CaCH$_3$ for which accurate frequency predictions are
available. Only upper limits have been obtained for these molecules.
}


\titlerunning{CaNC in space}
\authorrunning{J. Cernicharo et al.}

\maketitle

\section{Introduction}
Diatomic metal-bearing molecules were
predicted to be produced under thermodynamical chemical equilibrium near the photosphere
of carbon-rich evolved stars \citep{Tsuji1973}. 
Some of these species (NaCl, KCl, AlF, and NaF) were detected 30 years ago in the circumstellar
envelope (CSE) of the carbon-rich star envelope IRC+10216 by \citet{Cernicharo1987}. Shortly after this
detection, \citet{Guelin1986}
reported on the presence of a new free radical in IRC+10216. Although several silicon- and sulfur-bearing candidates were proposed
at that time, the line carrier was only identified 
from laboratory measurements to be magnesium isocyanide \citep{Kawaguchi1993}.
The emission of MgNC was mapped in IRC+10216 with the IRAM Plateau de Bure 
Interferometer by \citet{Guelin1993}. Unlike metal-halogen species
that are formed near the stellar photosphere, MgNC was found to arise from a thin shell 15''
in radius where many reactive species such as the carbon chain radicals C$_5$H, C$_6$H, C$_7$H, and
C$_8 $H, are also detected  \citep{Cernicharo1986a,Cernicharo1986b,Guelin1987,Cernicharo1987a,
Cernicharo1987b,Cernicharo1996,Guelin1997}.
Just one year after the identification of MgNC, sodium cyanide was also detected in IRC+10216
by \citet{Turner1994}. MgNC has been also found towards the more evolved carbon star CRL2688
\citep{Highberger2003}.

Since these early works, several more metal cyanides or isocyanides
have been found in CSEs. MgCN, AlNC, SiCN, SiNC, KCN, FeCN, and HMgNC have been identified
in IRC+10216 after their millimeter-wave rotational spectrum was characterized in the spectroscopic laboratory
\citep{Ziurys1995,Ziurys2002,Guelin2000,Guelin2004,Pulliam2010,Zack2011,Cabezas2013}.
Many others have been searched for in the same way without success. 

Metals such
as Na, K, Ca, Fe, and Cr and/or their cations are found in the gas phase in IRC+10216 
\citep{Mauron2010}, pointing towards a rich metal chemistry in the outer CSE.
Diatomic halogen metal-bearing species containing Al, Na, or K are stable closed-shell molecules 
that mostly form in the hot part of the envelope, close to the star \citep{Cernicharo1987}. However, the triatomic cyanides 
containing Mg, Si, or Fe are open -shell radicals that could react even at low temperature with neutral molecules or atoms and may be formed in the outer envelope.
The addition of an hydrogen atom to the radicals MgNC and SiCN yields stable closed-shell molecules that may be thought 
to be abundant in IRC+10216. HMgNC, the hydrogenated form of the abundant metal isocyanide MgNC, has been found
in IRC+10216 by \citet{Cabezas2013} with an abundance ratio N(MgNC)/N(HMgNC)$\simeq$20. The millimeter spectra of HSiCN and HSiNC have been accurately characterized 
in the laboratory by \citet{Sanz2002}. However, these species have so far not been detected in space. For HFeNC and
HFeCN only ab initio calculations are available, showing that the hydrogenated forms
of the cyanade and isocyanide compounds are also open-shell species \citep{Redondo2016}.

The chemistry of these metal-bearing isocyanides could be based 
on the reaction of the metal cations (Mg$^+$, Fe$^+$, K$^+$, Al$^+$, and Si$^+$) with other species
formed in the CSE \citep{Petrie1996,Dunbar2002}. 
The detection of additional metal-bearing 
species is therefore necessary to provide additional insights into the reactions leading 
to the formation of these molecules.

In this Letter we report on the first Ca-bearing species in space, calcium isocyanide (CaNC), and we perform detailed 
radiative transfer and chemical models to analyse the origin of the observed emission, and
the formation mechanisms of this species.

\section{Observations}
In the course of searches for new molecules we have covered a large portion of the 3, 2, 1, and 
0.8 mm spectrum of IRC+10216 with a high sensitivity
using the 30-m IRAM radio telescope. In the 3 mm window, the data 
acquired during the last 35 years cover the 80-116 GHz domain with very high 
sensitivity (1-3 mK). Examples of these data can be found in 
\citet{Agundez2008,Agundez2014}, \citet{Cernicharo1996}, \citet{Cernicharo2007,Cernicharo2008}, 
and references therein. Most of the 2 mm data come from the line survey of IRC\,+10216 
carried out by \citet{Cernicharo2000}, complemented with additional data obtained during the 
search for specific molecular species \citep[see, e.g.][]{Guelin2004,Agundez2008,Agundez2012,Fonfria2006}
and for the study of molecular emission variability with time \citep{Cernicharo2014,Pardo2018}. 
The sensitivity of the 2 mm observations varies between 1.5 and 10 mK. 
The 1 mm and 0.8 mm data come 
from observations carried out during the searches quoted above and from a line survey between 290 and 355 
GHz carried out using the new EMIR receivers in 2010 and 2011
with a sensitivity of 2-5 mK, depending on the atmospheric transmission. 
For frequencies above 250 GHz the spectrometers were two autocorrelators with 2 MHz of spectral 
resolution and 4 GHz of bandwidth. For all other observations the spectral resolution was 1 MHz, 
provided by filter banks or autocorrelators. 
Some recent observations used the new fast Fourier 
transform spectrometers, which cover a bandwidth of 2 $\times$ 16 GHz with a spectral resolution 
of $\sim$200 kHz. 

\begin{figure*}[]
\centering
\includegraphics[scale=0.77]{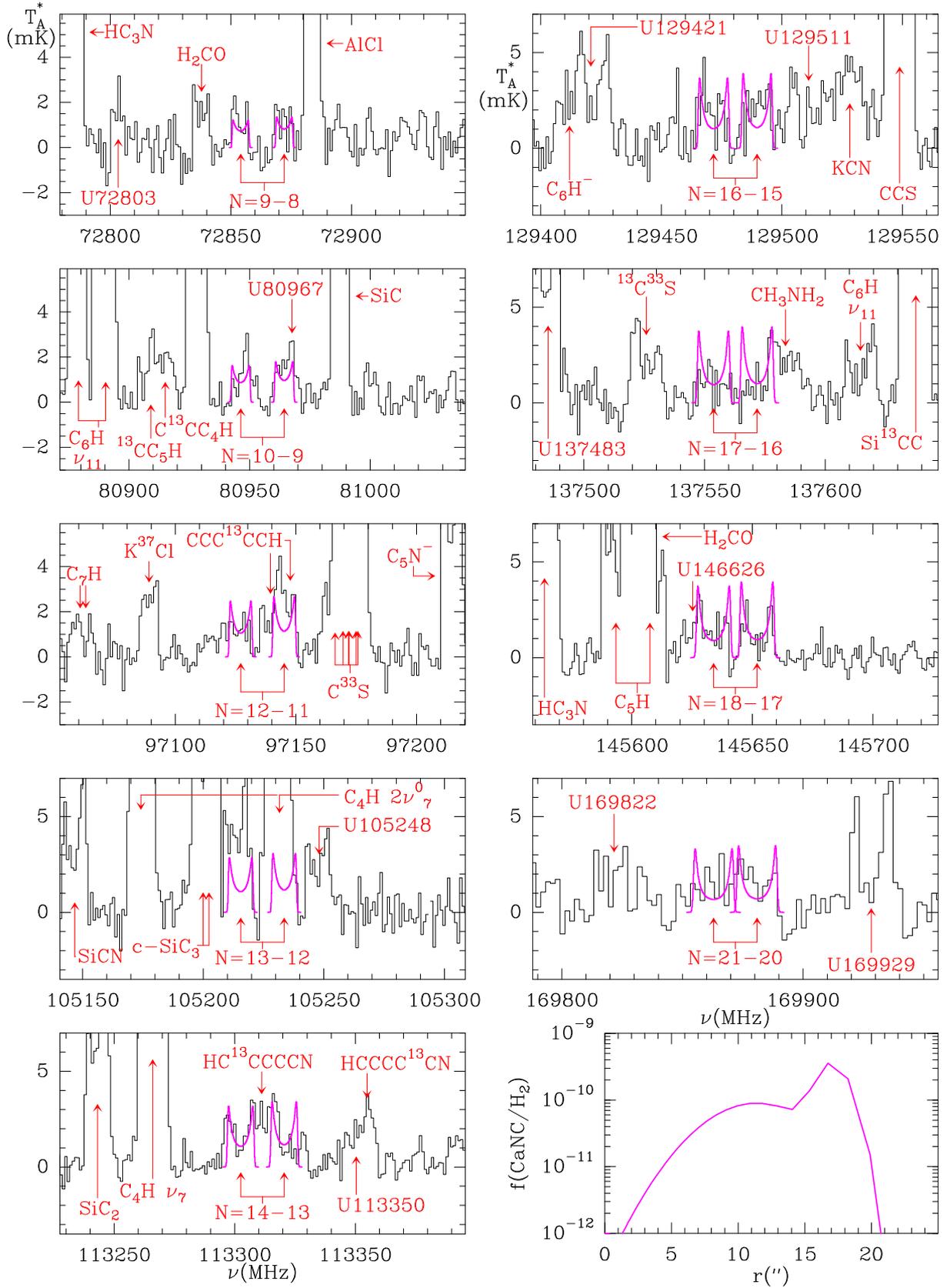} 
\caption{Results from the radiative transfer models, plotted in magenta, over the IRAM-30\,m spectrum of IRC+10216, the black histogram.
The vertical scale is the antenna temperature in mK, and the horizontal scale is the frequency in MHz.
Labels for the visible lines are plotted in red in each panel, where unidentified lines are labelled U lines.
The last panel in the bottom-right corner of the figure displays the radial abundance profile in magenta, which we used as input to 
obtain the synthetic spectra shown in previous panels. 
The vertical scale of this last panel is the fractional abundance of CaNC relative to H$_2$, and the horizontal scale 
is the radial distance to the star in arcseconds. 
}
\label{fig:canc}
\end{figure*}

The observing mode, in which we wobbled the secondary mirror by $\pm$90'' at a 
rate of 0.5 Hz, and the dry weather conditions (sky opacity at 225 GHz was below 0.1 in most 
observations) ensured flat baselines
and low system noise temperatures ($T_{\rm sys}$ $\simeq$ 100-400 K depending on the frequency).
This observing method, with the off-position located at 180$''$ 
from the star, provides reference data free from emission from all molecular species but 
CO \citep[see][]{Cernicharo2015}. The emission of all other molecular species is restricted 
to a region $\le$ 15-20$''$ from the star \citep[see, e.g.][]{Guelin1993,
Agundez2015,Agundez2017,Velilla2015,Quintana2016}.

The intensity scale, antenna temperature ($T_A^*$), was corrected for atmospheric absorption using the ATM 
package \citep{Cernicharo1985,Pardo2001}. 
Each frequency setup was observed for $\sim2$ h, with pointing and focus checks in between using strong nearby quasars. 
Pointing corrections were always within 2-3$''$. The 30m beam sizes are in the ranges $30''-21''$ at 3\,mm, $20''-17''$ at 2\,mm, 
and $12''-9''$ at 1\,mm.
The main-beam antenna temperature can be obtained by 
dividing $T_A^*$ by the main-beam efficiency of the telescope, which 
is 0.81, 0.59, and 0.35 at 86, 230, and 340 GHz, respectively. 
Calibration uncertainties for data covering such a long observing period have been adopted 
to be 10\%, 15\%, 20\%, and 30\% at 3, 2, 1, and 0.8 mm, respectively. 
Additional uncertainties 
could arise from the line intensity fluctuation with time induced by the variation of 
the stellar infrared flux, which has been discovered by \citet{Cernicharo2014} for a few molecules
and was recently
revisited for most molecules present in IRC+10216 by \citet{Pardo2018}. All 
data have been analyzed using the GILDAS package\footnote{http://www.iram.fr/IRAMFR/GILDAS}.

\section{Results}
The data revealed several hundreds of spectral lines that could not be assigned to any 
known molecular species collected in the public CDMS \citep{Muller2005} and JPL \citep{Pickett1998} 
spectral databases. Most of these lines show the 
characteristic U-shaped or slightly flatted profiles, with line widths of 29 km\,s$^{-1}$ , indicating
that they are formed in the external and intermediate layers of the envelope. Among these 
unknown features a series of doublets were found at 3 mm in our 
early observations prior to 2000. These doublets appear in harmonic relation with integer quantum numbers
for the rotational angular momentum. However, the sensitivity of the data at 2 mm was not enough to
detect additional doublets above 113 GHz. Only when the total set of data at 2 mm 
described in the previous section was available were lines up to $N_{up}$=21 detected.
The observed lines are shown in Figure\,\ref{fig:canc}. No lines above 170 GHz were found because the observed lines at 3 and 2 mm are weak.

The distortion constant of the carrier of the observed lines is
10 times higher than that of HC$_3$N, which has a rotational constant of 4549 MHz. However,
linear triatomic species such as MgNC, MgCN, AlCN, AlNC, SiCN, SiNC, FeCN, 
and FeNC have distortion constants between 2 and 6 kHz and rotational constants between $\simeq$4-6 GHz.
\citep{Kawaguchi1993,Robinson1997,Walker1999,Apponi2000,Flory2011,Zack2011}.
The
carrier of the doublets might therefore be a molecule containing an ionic bond, that is, a metal-bearing species. 
Moreover, the presence of doublets at constant separation of 18 MHz indicates that the carrier
could have a $^2\Sigma$ electronic ground state.
 
From the observed
rotational constant the isocyanide CaNC seems to be a good candidate as carrier of the observed lines. 
This species was implemented in the MADEX code \citep{Cernicharo2012} from a fit to the rotational lines observed in the
laboratory by \citet{Steimle1993} and \citet{Scurlock1994}, from which we derive
$B$=4048.73259(68) MHz, $D$=5.001(2) kHz,
$\gamma$=18.082(12) MHz plus some high-order distortion constants. Accurate frequency predictions can
be obtained with these constants
for all the rotational lines of CaNC within the frequency range of our observations.
Figure\,\ref{fig:canc} shows that all the observed doublets correspond to the rotational lines of CaNC. 
The high dipole moment of the molecule, 6.985 D \citep{Steimle1992}, has certainly helped in its detection.

The observed line intensities of CaNC are rather low, and at this level of sensitivity, the isotopologues
and vibrationally excited states of other molecules could introduce blending and confusion in
identifying weak features. 
Of the ten doublets reported in Figure\,\ref{fig:canc}, several are affected by blending. 
The $N$=9$\rightarrow$8, 10$\rightarrow$9,
12$\rightarrow$11, 14$\rightarrow$13,
16$\rightarrow$15, 17$\rightarrow$16, 18$\rightarrow$17, 
and 21$\rightarrow$20 clearly show the two doublets corresponding to the fine structure
of each rotational transitions. The worst case corresponds to the
$N$=13$\rightarrow$12 transition, for which the high-frequency component of the doublet
is fully blended with one of the lines of the doublet of the $N$=11$\rightarrow$10 transition
of C$_4$H in its 2$\nu_7$ vibrationally excited state. The low-frequency component of this
transition of CaNC is blended with C$_8$H. In some cases, one of the components of each doublet
is slightly blended left or right by weak features, as indicated in the figure. 
For example, the 19$\rightarrow$18 transition (not shown here)
is blended in the red side with SiC$_2$, HCCN, and with an unknown feature. The left
side of this CaNC doublet is blended with a feature from Si$_2$C \citep{Cernicharo2015b}. 
Nevertheless, the emission corresponding to CaNC is clearly visible between these
strong features. As commented previously, the expected intensities of the
rotational lines with $N>$21 are very low, and no lines are detected above 170 GHz
within the limit of sensitivity of our data.
We conclude that the number of observed transitions, with their two-line components, is enough to guarantee 
the detection of CaNC in IRC+10216.

\section{Discussion}\label{sec:rt}
We carried out a radiative transfer analysis aiming to derive the radial abundance profile of CaNC 
consistent with the emission lines observed.
We used the radiative transfer code MADEX \citep{Cernicharo2012}, which computes the population 
of the levels of a molecule and solves the radiative transfer problem for a spherical circumstellar envelope
divided into shells. The populations were calculated 
using the large velocity gradient (LVG) approximation \citep{Sobolev1960,Castor1970}
for each shell, starting from equilibrium conditions, and 
then the radiation field was computed, which is used to derive new populations, 
following the classical iterative approach until convergence was reached. The implementation of the 
LVG analysis was based on the formalism of \citet{Goldreich1974}.
The code then computed the emergent line profile by ray-tracing through the envelope, and by convolving the resulting 
profile with a Gaussian beam to reproduce the IRAM-30\,m main-beam response.

The physical structure of the CSE is essentially an updated revision of the model presented in \cite{Agundez2012},
where we implemented the mass-loss rate of the star and the kinetic temperature radial profile according to the most 
recent findings \citep{Guelin2018}.
There are no theoretical calculations for the collisional rates of CaNC, therefore we used the mass-corrected rates of 
the similar molecule MgNC with He \citep{Lique2013}, 
which is valid for the first 41 rotational levels (with $N$ between 0 and 20) in the temperature range [5--100]\,K. The
collisional rates for higher values of $N$ have been extrapolated from the calculated rates for the temperatures
of the layers introduced in the modelling.

The results from the radiative transfer models are presented in Figure\,\ref{fig:canc}.
Our model is able to reproduce the intensities and line profiles observed with the IRAM-30\,m telescope reasonably well.
Small discrepancies can be noted for the lines with poor signal-to-noise ratios, such as the $N$=21--20 doublet.
In addition to these small discrepancies and some blends in particular cases, we conclude that CaNC emission lines are detected 
in the millimeter wavelength spectrum of IRC+2016. 
According to our models, the CaNC abundance should be distributed in the intermediate envelope (from 2 to 10\arcsec) 
with an average value
of $\sim$3$\times$10$^{-11}$ (relative to H$_2$) and be significantly enhanced in a shell at $\sim$15--20\arcsec\ up to an average
value of $\sim$2$\times$10$^{-10}$.
The total column density of CaNC is 2$\times$10$^{11}$ cm$^{-2}$. 
Using the column density determined by \citet{Cabezas2013} for MgNC,
1.3$\times$10$^{13}$ cm$^{-2}$, the observed abundance ratio N(MgNC)/N(CaNC) is $\simeq$65. 

The derived column density of CaNC has to be considered as an upper limit because IR pumping through the low-energy $\nu_2$ bending
mode has not been considered in our calculations. \Citet{Nambu1997} have performed$\text{}$ $\text{ab initio}$ calculations for the potential energy
surface of CaNC and found that the bending frequency could be particularly low ($\simeq$27 cm$^{-1}$). Following
\citet{Agundez2006} and \citet{Agundez2017}, infrared pumping can significantly increase the population of intermediate and high-$N$ 
molecular levels with respect to the pure collisional excitation case. This pumping mechanism, which occurs through 
far-IR photons, can be very efficient indeed, even in the
external layers of the envelope where the lines of CaNC are formed. 

The density in the region where CaNC is predicted to be abundantly present, that is, at 15--20\,\arcsec\
\citep[or equivalently, $\sim$3$\times$10$^{16}$\,cm\ considering a distance of 123\,pc\ to the object;][]{Groenewegen2012},
should be close to $\sim$3$\times$10$^{4}$\,cm$^{-3}$, and kinetic temperatures should be close to 40\,K.
These values have been estimated by using the continuity equation with the mass-loss rate and also the temperature power law
derived by \cite{Guelin2018}.
Given the low densities in this region of the envelope plus the expected values for the critical densities of the transitions here presented
 ($\sim$10$^{4-5}$\,cm$^{-3}$), we cannot rule out uncertainties of a factor two or three in the fractional abundance derived
 due to radiative pumping effects and approximations considered.

\begin{figure}[]
\centering
\includegraphics[scale=0.52]{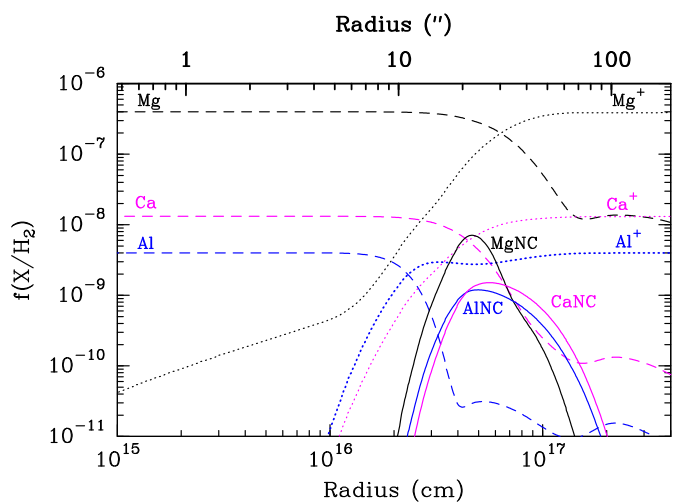} 
\caption{Abundances, relative to molecular hydrogen, for neutral and ionized Mg, Ca, and Al, together with
those of MgNC, AlNC, and CaNC resulting from the model discussed in the text. The lower X-axis shows the
distance to the star in centimeters, while the upper X-axis shows the distance in arcseconds assuming a distance
of 123 pc (see text). 
}
\label{figure2}
\end{figure}

The chemistry of Ca-bearing molecules, including CaNC, in circumstellar envelopes has been discussed by 
\citet{Petrie2004}. This author proposed that Ca-containing cyanides of the type Ca(C$_{2n+1}$N), with $n=0,1,2$, 
could be formed in a two-step process initiated by the radiative association of Ca$^+$ with long cyanopolyynes 
followed by the dissociative recombination with electrons of the Ca$^+$/NC$_{2n+1}$H complexes, which would 
fragment into the radicals Ca(CN), Ca(C$_3$N), and Ca(C$_5$N). This mechanism has been proposed to explain 
the presence of other metal cyanides such as MgNC, MgCN, HMgNC, NaCN, and AlNC in IRC\,+10216 
\citep{Petrie1996,Dunbar2002,Millar2008,Cabezas2013}. The plausibility of this formation mechanism 
is supported by the high rate coefficients calculated for the reactions of radiative association 
between the ionized metals Mg$^+$, Na$^+$, Al$^+$, and Ca$^+$ and cyanopolyynes larger than HC$_3$N 
\citep{Petrie1996,Dunbar2002} and by the observational evidence of the presence of neutral and ionized 
metals in the outer layers of IRC\,+10216 \citep{Mauron2010}. 

Chemical models have found that MgNC, MgCN, 
and HMgNC can indeed be formed with abundances in agreement with those derived from observations 
\citep{Millar2008,Cabezas2013}. To evaluate whether the presence of CaNC in IRC\,+10216 could 
similarly be explained by a similar mechanism, we have carried out chemical model calculations based 
on the recent model by \citet{Agundez2017} with a chemical network expanded to include metal chemistry 
as in \citet{Cabezas2013}. The chemical network involving Mg, Na, and Al has been taken from 
\citet{Dunbar2002}, while that involving calcium has been essentially taken from \citet{Petrie2004}. 
The results from our model are shown in Figure 2.

The observations of \citet{Mauron2010} indicate that gas-phase calcium atoms, which are mostly ionized 
in the outer shell of IRC\,+10216, have an abundance of $6.6\times10^{-9}$ relative to hydrogen nuclei. 
Interestingly, plugging this value as the initial abundance of atomic calcium in the chemical model results 
in a CaNC abundance of $\sim$10$^{-9}$ relative to H$_2$ (see Figure \ref{fig:canc} bottom right panel), 
somewhat above the value inferred from the observations and radiative transfer calculations. Uncertainties in the calculated 
abundance come mainly from the reaction rate coefficients, in particular, the branching ratios for the 
different fragmentations occurring in the dissociative recombination of 
Ca$^+$/NC$_{2n+1}$H complexes. The abundance derived from the observations probably has a large uncertainty 
due to the unknown effect of IR pumping through its bending mode on the excitation of the rotational levels 
of CaNC. When we take the different sources of uncertainty into account, the agreement between the calculated and 
observed abundance is quite satisfactory.

We have searched in our data for the Ca-bearing species for which accurate laboratory data are
available: CaF \citep{Anderson1994b}, CaCl \citep{Moller1982,Ernst1983,Domaille1977}, CaC \citep{Halfen2002}, 
CaCCH \citep{Anderson1995}, and 
CaCH$_3$ \citep{Anderson1996}. These species were implemented in MADEX \citep{Cernicharo2012}
and were searched for in our spectra of IRC+10216. However, none of then were detected within the sensitivity of the data. 
We derive 3$\sigma$ upper limits $\simeq$1-3$\times$10$^{11}$ cm$^{-2}$ for all them.
CaF, CaCl, and probably CaC might be formed in the 
innermost regions where
the other metal-halogen species have been found \citep{Cernicharo1987}. CaCCH and CaCH$_3$ might 
be formed in the external layers of the envelope following similar chemical paths to those producing
the metal-cyanide species. Except for MgCCH, which has been tentatively detected
in IRC+10216 by \citet{Agundez2014}, no detections of other metal-CCH species have been reported so far.
There are still many unknown features in our data that could correspond to metal-bearing species, in particular
those showing narrow features that are formed in the dust formation and growth zone, 1-20 stellar radii \citep{Cernicharo2013}.
While some of these features could correspond to vibrationally excited states of known molecules that are not yet well characterized
in the laboratory, many others certainly arise from the building blocks of the seeds of dust grains.
A significant laboratory effort has to be performed to find the carriers of these unknown lines.

\begin{acknowledgements}
The Spanish authors thank the Ministerio de Ciencia Innovaci\'on y Universidades for funding support from 
the CONSOLIDER-Ingenio program
``ASTROMOL'' CSD 2009-00038, AYA2012-32032, AYA2016-75066-C2-1-P. We also thank the
ERC for funding through grant ERC-2013-Syg-610256-NANOCOSMOS. 
MA thanks the Ministerio de Ciencia Innovaci\'on y Universidades for the Ram\'on y Cajal grant  RyC-2014-16277.
CB thanks the Ministerio de Ciencia Innovaci\'on y Universidades for the Juan de la Cierva grant FJCI-2016-27983.
LVP acknowledges support from the Swedish Research Council and ERC consolidator grant 614264.
We thank our referee, C. Gottlieb, for a critical reading of the manuscript.
\end{acknowledgements}

\end{document}